# Bidirectional optogenetic control of inhibitory neurons in freely-moving mice

Ori Noked*, Amir Levi*, Shirly Someck, Ortal Amber-Vitos, and Eran Stark

*These authors contributed equally to this work

**Abstract** – *Objective:* **Optogenetic manipulations of excitable cells enable activating or silencing specific types of neurons. By expressing two types of exogenectric proteins, a single neuron can be depolarized using light of one wavelength and hyperpolarized with another. However, routing two distinct wavelengths into the same brain locality typically requires bulky optics that cannot be implanted on the head of a freely-moving animal.** *Methods:* **We developed a lens-free approach for constructing dual-color head-mounted, fiber-based optical units: any two wavelengths can be combined.** *Results:* **Here, each unit was comprised of one 450 nm and one 638 nm laser diode, yielding light power of 0.4 mW and 8 mW at the end of a 50 micrometer multimode fiber. To create a multi-color/multi-site optoelectronic device, a four-shank silicon probe mounted on a microdrive was equipped with two dual-color and two single-color units, for a total weight under 3 g. Devices were implanted in mice expressing the blue-light sensitive cation channel ChR2 and the red-light sensitive chloride pump Jaws in parvalbumin-immunoreactive (PV) inhibitory neurons. The combination of dual-color units with recording electrodes was free from electromagnetic interference, and device heating was under 7 °C even after prolonged operation.** *Conclusion:* **Using these devices, the same cortical PV cell could be activated and silenced. This was achieved for multiple cells both in neocortex and hippocampus of freely-moving mice.** *Significance:* **This technology can be used for controlling spatially intermingled neurons that have distinct genetic profiles, and for controlling spike timing of cortical neurons during cognitive tasks.**

## I. INTRODUCTION

IN recent years, optogenetic control of excitable cells has become a leading method for manipulating and studying neural circuits [1], [2]. Optogenetic manipulations allow focusing an effect such as neuronal activation or silencing on specific cell types using appropriate genetic markers. For spatial focusing, genetic methods include localized viral injections [3] - [5], transgenic animals [6], and combinatorial combinations thereof [7], [8]. Spatial focusing using optical methods range in scale from whole regions using large-core optic fibers (100-200 μm core diameter, ~30,000 μm²; [9] - [13]); via small neuronal groups using small-core single or multi-mode fibers [14], [15]), small-core waveguides [16] - [18], or injectable LEDs (2,500 μm²; [19]); and down to individual cells, using μLEDs located within the brain tissue (150 μm²; [20]) or two-photon targeting [21] - [25]. For freely-moving animals, devices based on optic fibers allow flexibility in the choice of fiber diameter (and hence manipulated neuronal population size), light source (wavelength and opsin), and spatial relations between light emission points and electrodes.

Since distinct opsins are activated by different light wavelengths, focusing multiple wavelengths onto a single spatial location opens new possibilities for manipulating neural circuits. The value of dual-color control was recognized early on [8], [26], [5]. One application involves bidirectional control of a single population: silencing spontaneous neuronal activity with one wavelength (using a red-light sensitive neuronal silencer such as Halorhodopsin3 [8], Archeorhodopsin [27], or Jaws [28]), while generating synthetic spiking with a second wavelength (e.g. using the blue-light sensitive activator Channelrhodopsin-2, ChR2; [29]). A second application involves the independent control of spatially overlapping, genetically distinct cell types (such as PV and SOM cells) using optical depolarization tools sensitive to different wavelengths (such as ChR2 and the red-light sensitive opsin Chrimson; [30]). Other applications are clearly possible, such as the simultaneous yet independent control of distinct cellular compartments.

In recent years, several attempts were made to assemble dual-color optogenetic stimulation devices in freely-moving animals. Early solutions were "on-bench", involving physically large optical assemblies that couple light from multiple sources into a single fiber, and then routing that fiber to the animal [26], [13], [5], [31]. In these cases, light is transmitted to the animal via a rigid optic fiber (or fiber bundle) that restricts animal movement and has limited scalability. A complementary approach places all optical devices on the head of the animal [15], [20]. Using these head-mounted "diode-probe" devices, even small animals such as mice are free to perform complex

Manuscript received February 13, 2020; revised May 1, 2020; accepted June 6, 2020. This work was supported by a CRCNS grant (#2015577) from the United States-Israel Binational Science Foundation (BSF), Jerusalem, Israel, and the United States National Science Foundation (NSF); and by an ERC Starting Grant (#679253).

O.N., A.L., S.S. and E.S. are with the Sagol School of Neuroscience and with the Department of Physiology and Pharmacology, Sackler Medical School, Tel Aviv University, Tel Aviv 6997801, Israel. O.A.V. is with the Department of Physiology and Pharmacology, Sackler Medical School, Tel Aviv University, Tel Aviv 6997801, Israel.
Correspondence: eranstark@tauex.tau.ac.il.

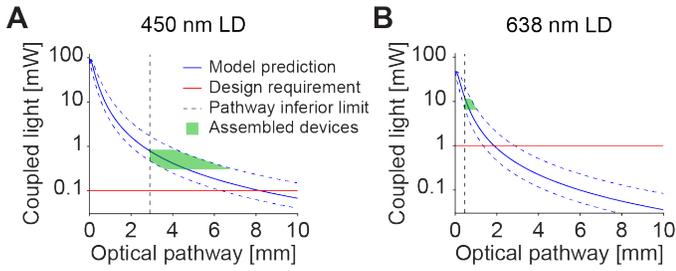

**Fig. 1. Predictive model.** Light output of each LD was modeled as a simple astigmatic Gaussian beam (solid blue curve). Numerical simulations were used to predict the longest optical pathway that will satisfy the design requirements (solid red line: 0.1 and 1 mW of 450 and 638 nm light, respectively). Parameters used for modeling are specified in Table 1. The longest optical pathways that satisfy these requirements are 8.19 and 1.87 mm for the 450 and 638 nm LDs, respectively. Based on the physical dimensions of the optical unit components (Fig. 2), the inferior limits of the pathways are shorter (2.90 and 0.46 mm for the 450 and 638 nm LDs, respectively). The maximal light output of the assembled devices was well above the design requirements (green background; range over n=8 devices, optical pathway was not measured).

behavioral tasks [32]. While it is possible to couple two different light sources (e.g. red and blue diodes) to different fibers within the same device to achieve dual-color control [15], such designs cannot produce reliable overlap of the illuminated targets. To achieve robust spatial overlap, multiple wavelengths must be routed into a single waveguide in a compact package. We have recently reported such an advance [16], [17], but that approach was tailored to specific wavelengths (405 and 638 nm), required custom-made parts (e.g. GRIN lenses), utilized a full nanofabrication facility, and was applied only to head-fixed anesthetized animals. Thus, to date, there is essentially no solution for generating dual-color light on-head for freely-moving animals.

Here, we present a novel design for a dual-color, head-mounted, fiber-optic based device combined with recording electrodes. The devices have low cost and light weight, are scalable and flexible, were tested extensively in vivo, and produce unique results.

## II. METHODS

### A. Design specifications and constraints

The device was designed for the stringent application of a freely-moving mouse, since in such animals genetic control allows high flexibility in opsin targeting. Each device was equipped with two dual-color and two single-color optical units, each coupled to an array of eight recording electrodes, and all integrated into a single device mounted on a movable microdrive (Fig. 5). The device had to fulfill five main requirements: (1) Each dual-color unit must have a compact size, quantified by a maximal weight of 700 mg to allow mounting multiple units on a freely-moving mouse (maximal weight of the entire implanted construct should be no more than 10-20% body weight, i.e. ~4.5 g for a 30 g mouse). (2) Each dual-color unit coupled to an electrode array (silicon probe shank) must have an intra-cortical insert diameter smaller than 70 μm. (3) The maximum operation current for each light source must be under 100 mA, to prevent diode heating and possible failure of thin light-weight 36 AWG wires. (4) The current-generating circuitry and light sources must be electromagnetically isolated, to minimize electromagnetic interference (EMI) and capacitive coupling between the current sink (diode) and the un-buffered electrophysiological records [17], [15]. (5) The light power at the tip of a 50 μm multimode fiber must be least 100 μW at a wavelength of 450 nm to allow activation of ChR2 [29], iC1C2 [33], and ACR [34]; and at least 1 mW at a wavelength of 638 nm to activate Halorhodopsin3 [8], Jaws [28], Arch [27], or Chrimson [30].

To reduce the number of moving components, assembly complexity, cost, and size, we decided to design a device without any lenses or active cooling. Since commercially-available laser diodes (LDs) have divergent angles (i.e. emit light at angles) larger than 10°, this decision required minimizing the optical pathway to achieve adequate coupling efficacy. The physical requirements were met using ø3.8 mm Transistor Outline (TO) can LDs. To maximize the dynamic range of the light output, we chose LDs that can operate within the maximal current limitation of 100 mA and have low threshold currents. The chosen diodes were PL450B (450 nm, Osram) and HL63603TG (638 nm, Ushio). Since future devices may require different operating wavelengths, additional commercially-available LDs that fulfill the same criteria were identified, including SLD3247VFR (405 nm, Sony) and PL520B2 (520 nm, Osram).

### B. Predictive modeling

To test the feasibility of several alternative optical designs, a predictive model was devised. Light output from each LD was modeled as an astigmatic Gaussian beam according to four parameters: LD wavelength ($\lambda$), light output power ($P$), and the horizontal and vertical divergent angles ($\theta_{FWHM,x}$, and $\theta_{FWHM,y}$, Table 1). Due to their asymmetric crystal, LDs generally emit light at an elliptical profile [35]. This profile can be described using the simple astigmatic Gaussian (SAG) model [36]. Like the Gaussian beam model, the SAG model is a solution of the paraxial Helmholtz equation. Unlike the Gaussian beam, the SAG beam has two q factors (1, 2), one for each major axis (x, y). The q factors have two alternative representations:

$$q_i = z - z_{0i} + i z_{R_i} \mid \mid \frac{1}{q_i} = \frac{1}{R_i(z)} - i \frac{\lambda}{\pi w_i(z)^2}, i = 1,2 \quad (1)$$

where $z_{0i}$ are the waist location of each axis; $z_{R_i}$ are the Rayleigh distance of each axis; $R_i(z)$ are the principal radii of curvature; $\lambda$ is the wavelength; and $w_i(z)$ are the principal

### TABLE I
SIMULATION PARAMETERS

| Parameter | Blue LD | Red LD |
|---|---|---|
| $\lambda$ | 450 nm | 638 nm |
| $P_{tot}$ (in air) | 88 mW | 45 mW |
| $\theta_{FWHM,1}$ (in air) | 11° / 7.5° / 4° | 13° / 8.5° / 3° |
| $\theta_{FWHM,2}$ (in air) | 25° / 21.5° / 18° | 13° / 18° / 23° |
| Alignment offset (each axis) | 5 μm / 10 μm / 15 μm | 5 μm / 10 μm / 15 μm |

When multiple values are listed, they refer to minimum, nominal and maximal values

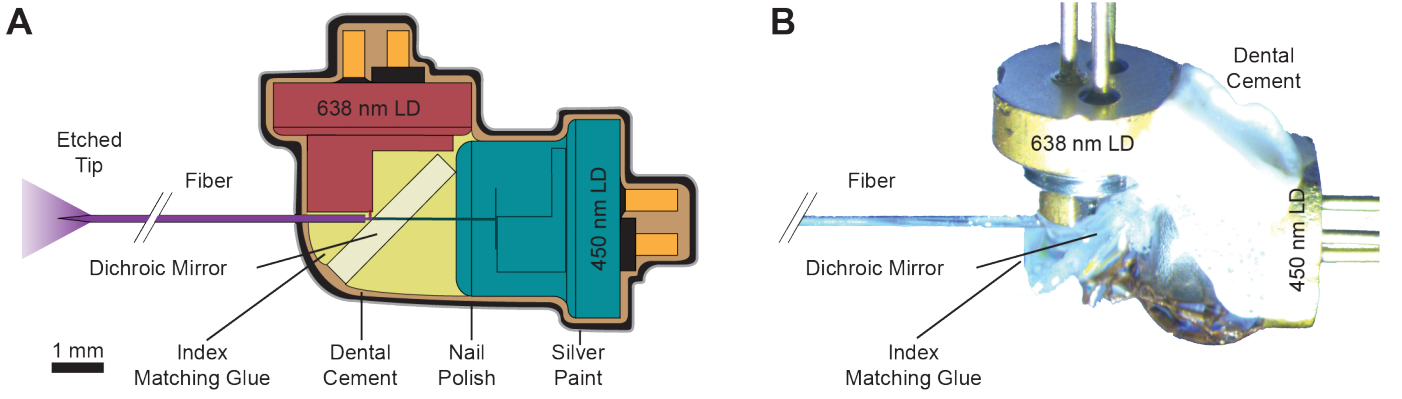

**A**

Etched Tip

638 nm LD

Fiber

450 nm LD

Dichroic Mirror

1 mm

Index Matching Glue

Dental Cement

Nail Polish

Silver Paint

**B**

Dental Cement

638 nm LD

Fiber

450 nm LD

Dichroic Mirror

Index Matching Glue

**Fig. 2. Optical unit design.** (A) Optical assembly design schematic. Light from the 638 nm LD is reflected at a dichroic mirror and then coupled to the fiber, while light from the 450 nm LD is transmitted via the mirror. The optical pathway is filled with index matching glue. The entire unit is encapsulated in three layers: dental cement (to add mechanical strength), nail polish (to block uncoupled light) and silver paint (to reduce electromagnetic interference). (B) A partially-assembled optical unit, photographed while placing the layer of dental cement. Scale is the same as in panel A.

semi-axes of the light spot ellipse. The light intensity of the SAG beam is:

$$I(x,y,z) = \frac{2P\sqrt{z_{R_1}z_{R_2}}}{\lambda|q_1q_2|} \exp\left(-\left(\frac{\sqrt{2}x}{w_1(z)}\right)^2\right) \exp\left(-\left(\frac{\sqrt{2}y}{w_2(z)}\right)^2\right) \quad (2)$$

The laws for propagation and refraction can be described using the ABCD law for both factors: given a beam with $q_i$ factor transitioning through an optical system with an ABCD matrix $M = \begin{pmatrix} A & B \\ C & D \end{pmatrix}$, the output q factor $q_i'$ can be calculated by:

$$q_i' = \frac{Aq_i+B}{Cq_i+D}, \quad (3)$$

where the ABCD matrices for propagation and refraction are $M_p$ and $M_r$ respectively:

$$M_p = \begin{pmatrix} 1 & d \\ 0 & 1 \end{pmatrix}, \quad M_r = \begin{pmatrix} 1 & 0 \\ 0 & \frac{n_{in}}{n_{out}} \end{pmatrix}. \quad (4)$$

The Rayleigh distance of each axis is calculated based on the specifications of each LD: the divergent angle $\theta_{FWHM,i}$, the wavelength $\lambda$, and $M^2$ (we used 1.3 throughout):

$$z_{R_i} = \frac{4\lambda M^2}{\left(\theta_{FWHM,i}*1.699\right)^2*\pi} \quad (5)$$

Fresnel losses in the transition between media were calculated according to

$$T = \frac{4n_1n_2}{(n_1+n_2)^2} \quad (6)$$

All optical pathways were simulated with an index matching medium with refractive index (RI) of 1.54 and RIs of 1.5 for the LD cap and the fiber. All calculations assumed near perfect alignment between the LDs and the optic fiber (5-15 μm offset in each axis; Table 1). The model predicted that a driving current of 100 mA, a sufficient amount of light (100 μW) will be coupled from the 450 nm LD to a ø50 μm optic fiber, even when the fiber is positioned 8.6 mm away from the LD. The model also predicted that if the 638 nm LD is positioned 2.0 mm away from the optic fiber, 1 mW would be coupled (Fig. 1).

Given the design requirements and the size of each LD (diameter of several mm), the results of the predictive model indicated that a butt-coupling solution [15] cannot be extrapolated to the dual-source case simply by angling one or both of the diodes. Thus, we decided to set the two LDs at a 90° angle relative to each other, and used a dichroic mirror to co-align their beams into the same fiber (Fig. 2). A dichroic mirror is ideal for this purpose since it has low optical power losses, thickness of one mm or less, and can be cut to a small size (several mm²). UV-curable clear glue (NOA68, Norland; RI=1.54) was used to fill the optical pathway. The glue allows optimization of coupling efficiency before curing, adds mechanical strength, and acts as an index matching medium that reduces reflective (Fresnel) losses. In this configuration, the minimum feasible optical pathways due to mechanical considerations are 2.9 mm for the 450 nm LD and 0.46 mm for the 638 nm. Assuming a perfect dichroic mirror and optimal coupling, the upper limits of coupled light power were estimated to be 786 μW and 12.2 mW for the 450 nm and 638 nm LDs, respectively (Fig. 1).

### C. Assembly process

To minimize the volume of the displaced tissue as well as tissue damage, we used a 50 μm core optic fiber (FG050UGA, Thorlabs), etched to a cone (length, 175 μm) in a mineral oil/hydrofluoric acid interface [15] at the distal (tissue) end. The fibers were stripped from the acrylate jacket and cut to a length of 30 mm. To effectively couple light from the LDs, the fibers were cleaved straight at the proximal (LD) side. To minimize the optical pathway and total device size, we removed the cap of the TO can from the 638 nm LD using a dedicated tool (WR1, Thorlabs). Since UV-curable glue applied to the 450 nm LD without a cap caused diode failure, we used the 450 nm LD with an intact TO can. Both LDs were driven by a precision current source with a 100 mA square wave, 3 s period, and 20% duty cycle (i.e. 600 ms continuous-wave [CW] on, to prevent over-heating), for 30 cycles at a time. Light power was measured at the 30th cycle. To manipulate the LDs in free space, we used a set of 3-axis ½" linear micromanipulators (M-460A-XYZ, Newport; Fig. 3A) connected to clamps (PC2/M, Thorlabs) that held custom 3D printed holders for the LD sockets (Fig. 3C). To direct the beams from two LDs into the same fiber, we used a 0.5 mm thick, 550 nm shortpass dichroic mirror (T550SP, Chroma). These commercially-available mirrors were cut to 6 − 9 mm squares (OPTEC Optical Components).

Assembly began by gluing the etched optic fiber to the dichroic mirror. The mirror was fixed in place on a rotation stage (RP01/M, Thorlabs) using a custom miniature clamp (Fig. 3E) and rotated to a 45° angle relative to the breadboard (Fig. 3B). To protect the mechanical assembly (breadboard and rotating stage) from drops of glue, we placed a piece of non-stick vinyl glove between the clamp and the mirror. The optic fiber was held by a 3D-printed plastic holder consisting of a vertical black sheet that prevented direct illumination of the light power sensor (Fig. 3D). The fiber itself was passed through a ø0.3 mm pinhole, and residual light was blocked by a small piece of putty. Another XYZ micromanipulator connected to a clamp was used to position the fiber holder. The fiber holder was aligned with the optic breadboard and, using the XYZ micromanipulator, the fiber was placed at a touching distance from the mirror surface. The fiber was then glued to the dichroic mirror using UV curable glue (NOA68, Norland; cured using CS2010, Thorlabs).

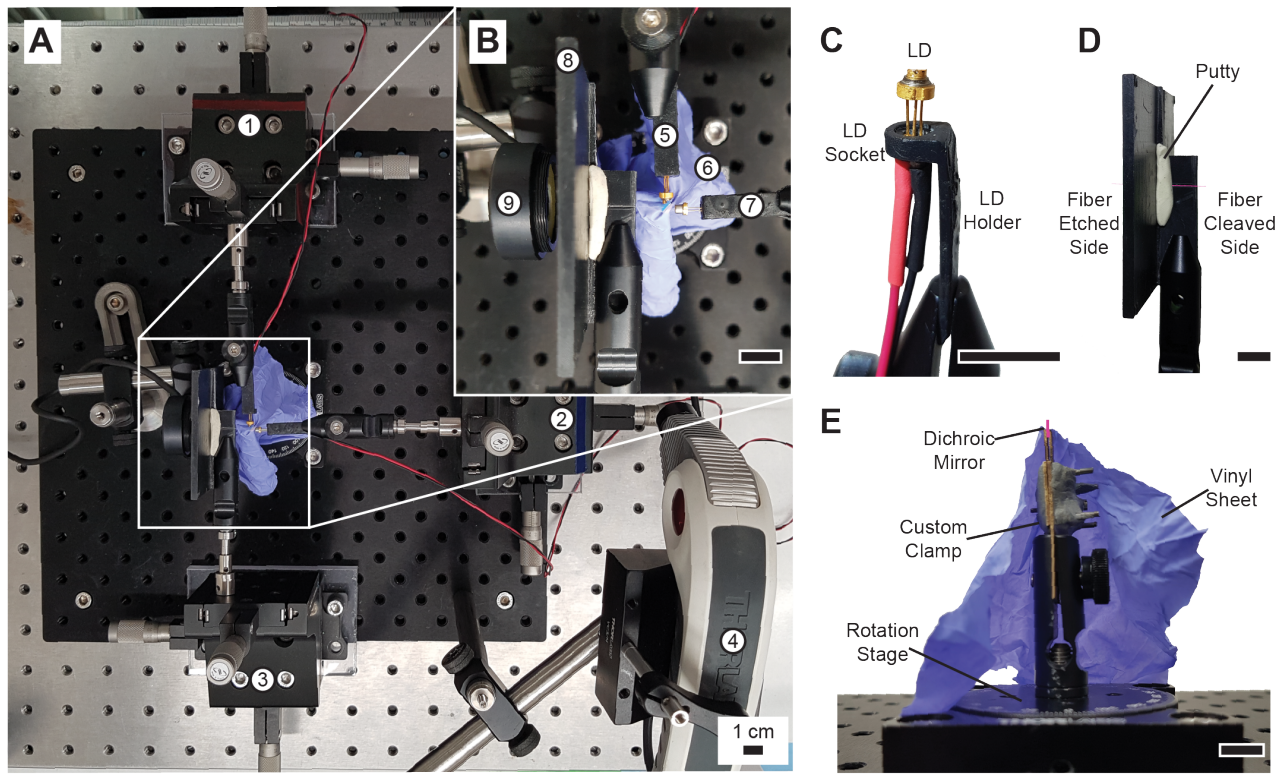

**Fig. 3. Optical unit assembly setup.** (**A, B**) Optical breadboard (top view). The dichroic mirror is held at the center of the breadboard using a custom-made clamp (**6**) while three micromanipulators (**1, 2, 3**) are used to position the two LDs and the optic fiber. 3D printed parts are used to join each LD with its clamp (LD holders, **5, 7**) and to join the optic fiber with its clamp (fiber holder, **8**). Coupled light is monitored using a photosensor (**9**). Once optimal placement is achieved, index matching glue is cured using the UV light curing system (**4**). (**C**) LD close-up (lateral view). The L-shaped LD holder provides a stable mechanical interface as well as an electrical connection via an LD socket. (**D**) Optic fiber holder close-up (top view). The fiber is highlighted in pink. The fiber holder consists of a fiber tray connected to a light blocking sheet. The fiber (OD, ø0.15 mm) is placed on the tray and passes through a ø0.3 mm pinhole, while the sheet blocks uncoupled light from reaching the photosensor. After placing the fiber, a piece of putty is used to block the rest of pinhole and secure the fiber in place. (**E**) Dichroic mirror clamp close-up. The mirror is highlighted in pink. The dichroic mirror is held using a miniature clamp placed on a rotary stage. This allows positioning the mirror at a 45° angle relative to the LDs. A vinyl sheet protects the clamp and rotary stage from glue drops. Scale bar is 1 cm in all panels.

Next, light from the 638 nm LD was coupled into the fiber. A light power sensor (S130C, Thorlabs) connected to a photometer unit (PM100D, Thorlabs) was placed at the distal tip of the fiber to measure the coupled light. The 638 nm LD was placed at 90° from the fiber, at a touching distance from the proximal tip of the fiber, and aligned with the breadboard. Several drops of UV curable glue were used to fill the gap between the LD and the fiber tip. The LD was operated, and the

XYZ micromanipulator was used for modifying LD location and maximizing light output. Once a reading of at least 7 mW was achieved (15% coupling efficiency), UV curing began. To prevent mechanical movements caused by the curing process due to anisotropic glue shrinkage, curing was done while gradually increasing UV intensity: sequential 2 min exposures to 50, 100, and 150 mW/cm², followed by final curing by 10 min at 175 mW/cm². During the curing process, light output at the distal tip of the fiber was monitored. Any reduction in light output was interpreted as shrinking-induced mechanical movement and corrected by repositioning the LD with the micromanipulator prior to final curing. The 450 nm LD was positioned at the other side of dichroic mirror, and a similar coupling process (using the micromanipulator and photometer) was carried out until achieving power of at least 400 µW (0.5% coupling efficiency). The reduced coupling efficiency of the blue LD compared to the red LD (0.5% vs. 15%) is expected from the different lengths of the optical pathways (Fig. 1). The 450 nm LD was glued to the dichroic mirror with a drop of dental cement (to reduce glue shrinkage affects) before the same UV curing process used for the red LD was carried out.

To minimize the final size of the optical assembly, we cut the LDs legs to 1-2 mm and soldered 36 AWG wires (36744MHW, Phoenix Wire Inc.) to each LD anode and cathode. To achieve long-term mechanical stability and electrically isolate the LDs, the entire device was coated with a thin layer of dental cement (Grip Cement powder 675571 and liquid 675572, Dentsply). Uncoupled light was blocked from exiting the optical area by painting the device with black nail polish (486, GA-DE). To isolate the electrical fields generated by the voltage drops across the LDs from the recording electrodes and minimize EMI, the entire device was encapsulated within a Faraday cage by coating it with a 60% silver in water solution (SP-60+, M.E. Taylor Engineering Inc.), shorted to ground using another 36 AWG wire.

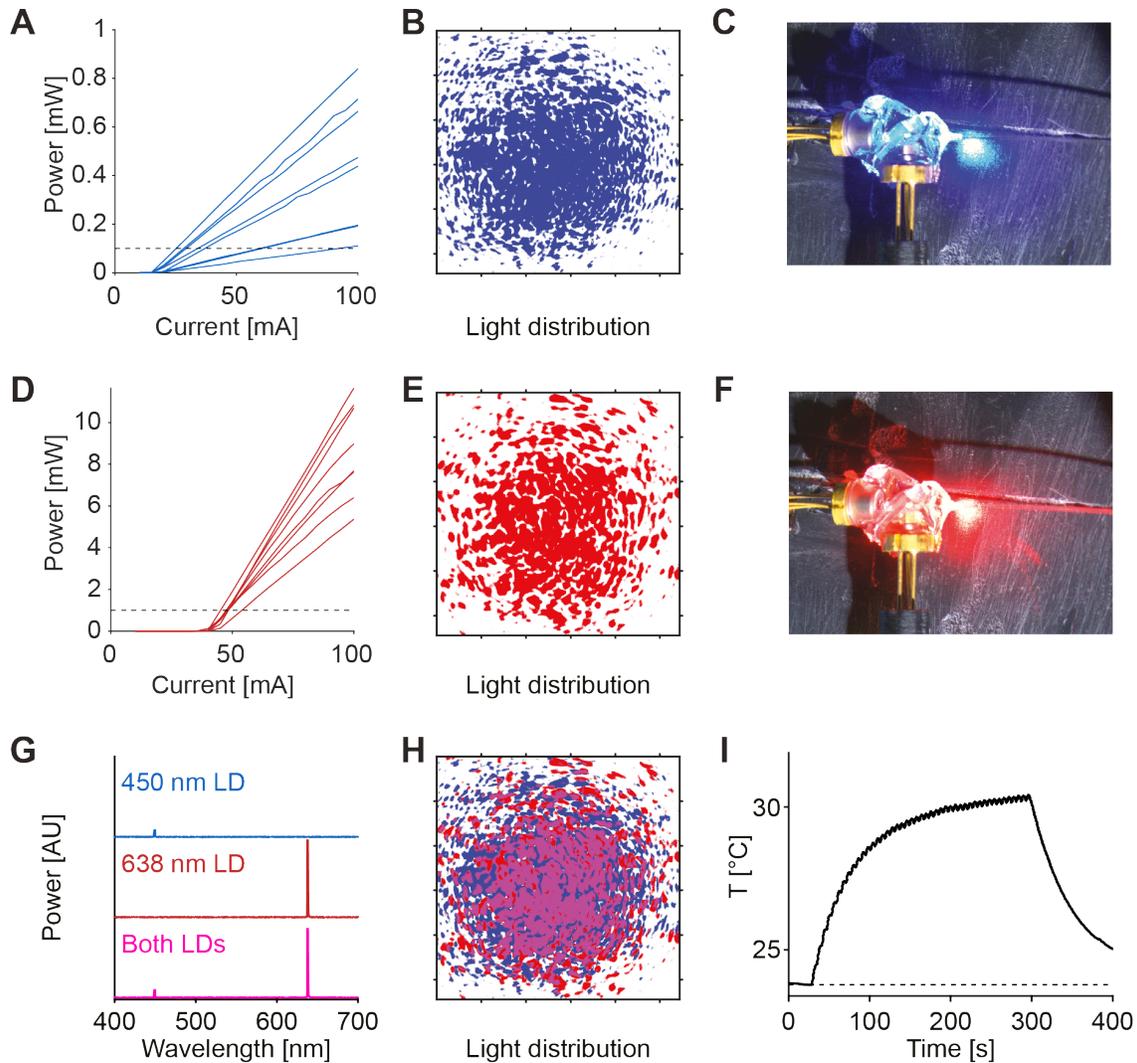

**Fig. 4. Optical unit performance assessment. (A, D)** Power-current (P-I) curves of the assembled units (blue (top) and red (bottom) LDs). Optical power at the etched tip of each fiber as a function of input current. Each line represents a different unit. All optical units satisfy the design requirements (for both wavelengths) above 50 mA. **(B, E)** Illumination profiles of a single optical unit. Illuminated regions are defined as different from the background distribution (higher than 3 SDs above the mean noise). Spatial overlap (defined as intersection/union) is 46%. **(G)** Emission spectrum of both LDs during simultaneous operation. No spectral interference is observed. **(H)** Overlapping illumination cross-section profile of a single optical unit: blue – 450 nm, red – 638 nm, purple – both wavelengths. **(I)** Temperature of a single optical unit under prolonged high load. Test conditions were full-power (100 mA) activation of both LDs: square wave, cycle duration of 6 s, and duty cycle of 10% (600 ms pulses), for 45 cycles. Temperature reached after 4.5 min was 31 °C (increase of 7 °C compared to initial condition of 24 °C).

### D. Preparation of optoelectronic probes

Multi-site optoelectronic probes were prepared as previously described [15]. First, a four-shank/32-channel silicon probe (Buzsaki32, NeuroNexus) was mounted on a microdrive. After cleaning in 2.5% Contrad 70 (Decon Laboratories; 60 min in 60°C), the impedance of each recording site, measured in 0.9% NaCl at 1 kHz (NanoZ, White Matter), was 1.14±0.11 MΩ (mean±SD over a total of n = 96 recording sites in three probes). Second, two dual-color (blue/red LDs) units and two single-color (blue LED) units were prepared for each probe. The impedance between the anode-cathode of each light source (LD/LED) and the silver-paint shield, measured at 1 kHz (Impedance Conditioning Module, FHC), was 20.4±9.9 MΩ (n=18 light sources utilized in 3 probes). Third, an optical unit was mounted on every shank: in each probe, shanks 1 and 4 received dual-color units, and shanks 2 and 3 received single-color units (Fig. 5). The weight of an implantation-ready device was 2.77 g, and implantations resulted in a weight increase of 5.2±0.7 g (equivalent to an increase of 16±3% from the baseline

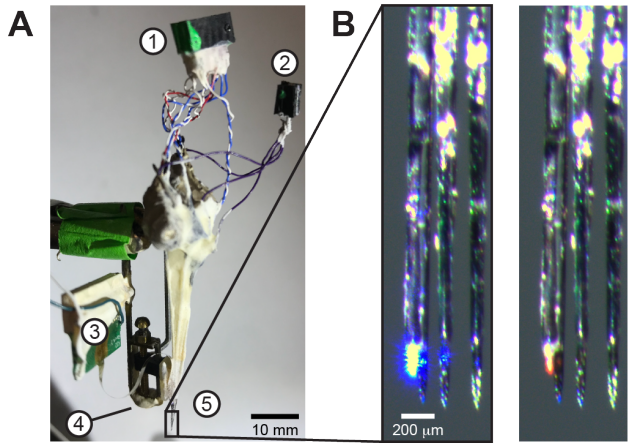

**Fig 5. Dual-color probe. (A)** Photo of an implantation-ready dual-color probe (**5**) mounted on a microdrive (**4**). Diode current connector (**1**), ground shield connector (**2**), neuronal recording connector (**3**). **(B)** Magnified photos three shanks during illumination of the blue (left) and red (right) LDs mounted as a dual-laser unit on shank 1. In each probe, shanks 1 and 4 received dual-color units, and shanks 2 and 3 received single-color units.

weight; n=3 mice weighing 35.9, 31.2 and 30.9 g before implantation).

### E. Animals and surgery

To express two opsins, ChR2 [37] and Jaws [28], in parvalbumin-immunoreactive (PV) inhibitory interneurons, three freely-moving male mice were used, all heterozygous for PV-Cre (aged 20, 14 and 32 weeks at the time of implantation). All animal handling procedures were in accordance with Directive 2010/63/EU of the European Parliament, complied with Israeli Animal Welfare Law (1994), and approved by the Tel Aviv University Animal Care and Facilities committee (01-16-051).

The first mouse was an offspring of a homozygous PV-Cre male (#008069, Jackson Labs) and an FVB/NJ female (#001800, Jackson Labs). This mouse was injected with a mix of two viral vectors, hSyn-Flex-Jaws (rAAV8/hSyn-Flex-Jaws; viral titer estimated at $3.2 \times 10^{12}$ IU/mL; University of North Carolina [UNC] viral core facility, courtesy of E.S. Boyden) and DIO-hChR2 (rAAV5/EF1a-DIO-hChR2(H134R)-eYFP; $3.2 \times 10^{12}$ IU/mL; courtesy of K. Deisseroth). The viral mix was injected stereotactically (Kopf) into neocortex and hippocampus at 8 different depths (AP -1.6, ML 1.1, DV 0.4 to 1.8 at 0.2 mm increments; 50 nl/site; Nanoject III, Drummond).

To increase the probability of dual-opsin (ChR2 and Jaws) expression by the same PV cell, a second procedure was used with two other mice, offspring of homozygous PV-Cre males and females homozygous for the Rosa-CAG-LSL-ChR2(H134R)-EYFP-WPRE conditional allele (#012569, Jackson Labs). Thus, ChR2 was innately expressed in PV cells throughout the brain of these mice [6]. To additionally express Jaws in the ChR2-expressing PV cells, we injected the hSyn-Flex-Jaws viral vector in the above-described coordinates (200 nl/site).

Four to eight weeks after injection, all animals were implanted with the multi-site/multi-color optoelectronic probes following previously-described procedures [15]. Briefly, each animal was placed under isoflurane anesthesia (1%). After preparing the skin for surgery (ethanol 70% and povidone-iodine 10%), local anesthesia (lidocaine-epinephrine 0.5 mg/kg SC) was applied. A midsagittal cut was made, the skull was cleaned (hydrogen peroxide, 3%), and four screws (000-120x1/16; Small Parts) were implanted in the skull. Two screws implanted above the frontal lobe were used for structural integrity of the copper Faraday cage, and two screws implanted above the cerebellum also served as ground and reference leads. After making a craniotomy and excising the dura mater, the diode probe, mounted on a microdrive, was inserted to a depth of 200-400 µm, and a silicon mixture (3-4680, Dow Corning) was applied to cover the brain and prevent infection. A copper mesh cage was built around the probe for mechanical and electromagnetic shielding, and the recording and stimulating connectors were cemented to inside of the mesh wall. Animals were treated with postoperative antibiotics (7.5 mg/kg; Enrofloxacin) and analgesics (0.05 mg/kg; Buprenorphine).

### F. Recording procedures

Neural activity was filtered, amplified, multiplexed, and digitized on the headstage (0.1–7,500 Hz, x192; 16 bit, 20 kHz; RHD2132, Intan Technologies). Animals were equipped with a 3-axis accelerometer (ADXL-330 or ADXL-335, Analog Devices) for monitoring head-movements. Each diode was driven by a separate channel of a custom-made 16-channel linear current source [15], controlled by a programmable DSP running at 24.414 kHz (RX8; Tucker-Davis Technologies) via ActiveX and MATLAB (The MathWorks, Natick, MA). Voltages proportional to the applied currents were recorded by separate channels of the data acquisition system.

Recordings were carried out in the home cage during spontaneous behavior. After each session, the probe was either left in place or advanced in 35-70 µm steps and the brain was allowed to settle overnight. At each location in the brain, neuronal activity was inspected for spontaneous spiking activity, and if encountered, a full recording session (>3 hours) was conducted. The session consisted of a baseline period of at least 15 min, followed by photostimulation. Initial stimulation was performed by each diode separately using 50-70 ms (blue) or 100-300 ms (red) light pulses at the minimal intensity that evoked an effect detectable by visual inspection during the experiment ("spiking threshold", defined using the driving current $I_{max}$) and at multiple intensities above and below that level (log-spaced multiples of $I_{max}$; typically 0.5, 1, 2, and 4).

### G. Electrophysiological data analysis

For offline analysis, spike waveforms were extracted from the wide-band recorded signals and sorted into individual units [38]. Briefly, waveforms were linearly detrended, projected onto a common basis obtained by principal component analysis of the data, and sorted automatically [39] followed by manual adjustment. Only well-isolated units (amplitude >50 µV; L-ratio <0.05 [40]; interspike interval index <0.2 [41]), were used. Subsequently, units were tagged as excitatory/inhibitory based on peaks/troughs in the short-time (±5 ms) pair-wise cross correlation (p<0.001, convolution test; [42]); and/or as PV-cells, based on a rate increase during 50 ms blue light pulses (p<0.01, Poisson test) or a rate decrease during 200 ms light pulses (p<0.01, Poisson test). Untagged units were classified as a putative PYR or INT based on a Gaussian-mixture model (p<0.05; [38]). We recorded a total of 230 well-isolated cells from the neocortex and CA1 of the three mice. Of these, 147 were PYR and 83 were INT.

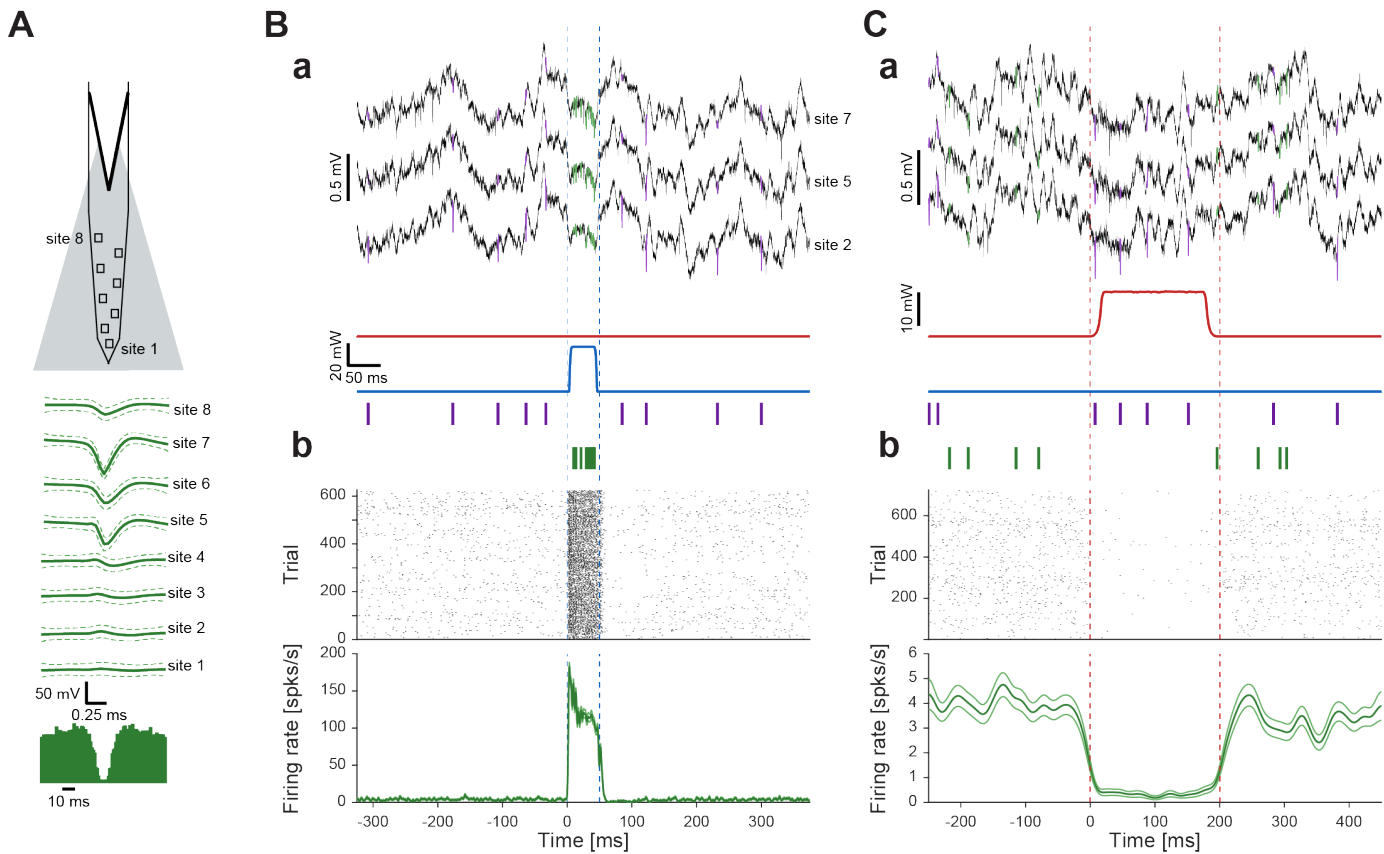

Fig. 6. Dual-color units allow bidirectional control of PV cells in freely-moving mice. (A) Top: schematics of a single shank of the dual-color probe with site numbering. Bottom: wideband (0.1–7,500 Hz) spike waveforms and autocorrelation histogram of a bidirectional controlled PV interneuron. (Ba) Wide-band traces and spikes of a well-isolated unit in neocortical recordings. Top three traces are neuronal recordings; two traces below indicate current applied to the LDs on the same shank. During the application of blue light pulses driven by a current of 29 mA, the unit increases its spiking activity. Spikes timing of the dual activated unit (green vertical lines) and a putative mono-synaptically connected PYR (purple vertical lines; see also Fig. 8Ba) during a single trial, are shown below the current traces. (Bb) Raster plot (top) and PSTH (bottom) during blue light pulses for the example unit. (Ca, Cb) Same as B, during red light pulses (driving current, 43 mA). The unit decreases its spiking activity.

## III. RESULTS

### A. Bench testing of dual-color optical units

Under the proposed assembly process (Fig. 3), eight dual-colored units were constructed (Fig. 2, Fig. 4C, 4F). The average unit weighed 630±55 mg (mean±SD). The optical evaluation of the assemblies consisted of input current vs. coupled light (P-I) curve for each unit, beam profile at the fiber tip, and emission spectra (Fig. 4). The light power measured at the tip of the fiber at a driving current of 100 mA was 454±271 µW and 8.65±2.26 mW for the 450 nm and 638 nm wavelengths, respectively (Fig. 4A, 4D). These values were 58% and 71% of those expected for the minimum feasible optical pathways (which are 786 µW and 12.2 mW, respectively; black dashed lines in Fig. 1). This is consistent with the fact that in practice, longer pathways were often used. Thus, light output power was well above the specifications set at the design stage (100 µW and 1 mW; horizontal dashed lines in Fig. 4A, 4D), allowing the usage of lower driving currents in animal experiments.

To allow independent dual-color control of the same neuron high light power does not suffice, and two additional requirements must be met. First, the cell body (~10 µm diameter) must be exposed to both wavelengths. Second, the two wavelengths must not interact optically, leaving all interactions to the electrochemical processes within the cell. The illumination profiles (measured using BeamPro 3.0 beam profiler, Edmund Optics) of the two LDs consisted of many speckles (Fig. 4B, 4E), as expected for multimode fibers. Nevertheless, the illumination profiles of two LDs coupled to the same fiber overlapped by 46% (Fig. 4H). This value is a lower-bound estimate of the overlap in the brain, since light scattering in tissue is expected to increase the overlap. In contrast, the spectra at the end of the fiber (measured using CCS100 spectrometer, Thorlabs) did not overlap, and there was no spectral interference during dual-color illumination (Fig. 4G). Thus, the optical characteristics of an isolated dual-color unit are suitable for bidirectional optogenetic control.

The inclusion of multiple dual-color units in an optoelectronic probe mounted on the head of a freely-moving animal raises two additional concerns: electromagnetic interference (EMI) with the electrophysiological signals and heating of the optical devices. The probability to obtain EMI was minimized by isolating the LDs, enclosing the entire optical unit within a Faraday cage (60% silver paint), and grounding the cage (Fig. 5A, 2). The isolation between the anode-cathode and the cage was 14.9±5.3 MΩ (n=8 dual-color units), indicating that EMI would be negligible [15]. Quantifying thermal effects is important since LD output power is inversely related to diode temperature, and since excessive device heating may affect brain temperature. Thermal effects were measured using a resistance temperature detector (NB-PTCO-162; TE Connectivity) glued onto the optical unit, and activating the unit periodically over a prolonged period. In room temperature, the peak temperature reached was under 31°C (Fig. 4I), lower than mammalian body temperature. Thus, the electrical and thermal characteristics of an isolated dual-color unit are compatible with simultaneous electrophysiological recordings.

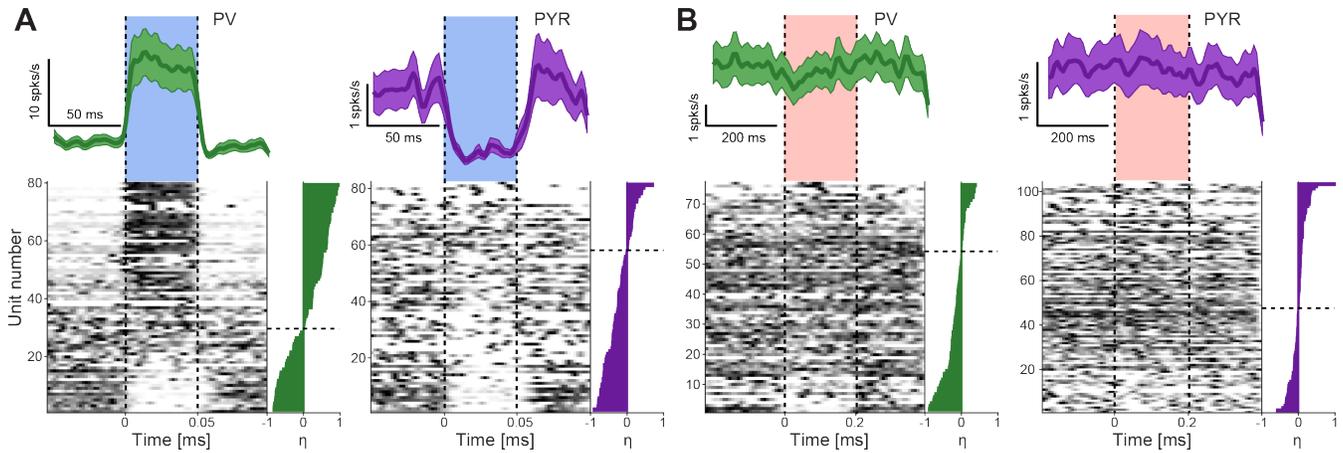

**Fig. 7. Population responses to blue and red light pulses** (**A**) PSTHs of putative and optically verified PV interneurons (left) and putative PYR (right) during blue light pulses. *Top*: Population response (mean ±SD). *Bottom*: Each row is a PSTH of a single unit, with higher firing rates shown in black. The units are sorted, from bottom to top, according to $\eta = \frac{r_{in} - r_{out}}{r_{in} + r_{out}}$, where $r_{in}$ ($r_{out}$) is the firing rate during the pulse (baseline). (**B**) Same as A for red light pulses.

### B. Bidirectional control of individual inhibitory cells in freely-moving mice

To determine whether the dual-color units can both activate and silence neurons in freely-moving mice, we carried out recordings with mice in which the blue-light sensitive cation channel ChR2 and the red-light sensitive chloride pump Jaws were both expressed in the same cell type. To minimize network-induced effects generated by trans-synaptic activation, we chose an inhibitory cell type: parvalbumin-immunoreactive neurons (PV cells). Thus, we implanted dual-color probes (Fig. 5) in the cortices of three freely-moving mice expressing both ChR2 and Jaws in PV cells, and monitored spiking and LFP activity (Fig. 6Ba, Ca).

Of 230 well isolated units recorded, some had very low spontaneous (baseline) firing rates (under 0.01 spikes per s). The probability that such a unit will fire during a 50 ms period is small enough that at least 6000 pulses are needed in order to determine whether firing rate decreased during light. In all recorded sessions, smaller numbers of equal intensity pulses (blue or red) were used. Thus, units were deemed valid for statistical analyses (Fig. 7) only if their spontaneous spike rate was above 0.01 spks/s. A total of 162 valid units (80 INT, 82

PYR) were used for the blue light analyses, and 181 (77 INT, 104 PYR) for the red light analyses. During blue LD illumination, some well-isolated units increased their spike rate (36 of 80 INT recorded in neocortex and CA1 of three mice; 45%; see Fig. 6B for an example neocortical unit and Fig. 7A for the population). The effect size for these units was $\eta = 0.61±0.04$ (mean ± standard error of the mean). During red LD illumination via the same fiber, some well-isolated units decreased their spike rate (37/77, 48%; $\eta = -0.37±0.04$; Fig. 7B, Fig. 6C). Thus, blue light induced stronger effects than red light ($p << 0.001$, Mann-Whitney $U$-test), possibly due to limited application of red light or to limited expression of the red-light sensitive opsin. Regardless, both opsins were effectively activated in the same animals.

A second order effect was observed in simultaneously-recorded PYR. During blue LD illumination, some PYR (18/82, 22%; Fig. 7A, right) decreased their spike rate ($\eta = -0.64±0.05$). These numbers are higher than expected by chance (1%; p<0.001, exact Binomial test). During red LD illumination some PYR (6/104, 6%; Fig. 7B, right) increased their spike rate ($\eta = 0.34±0.14$). No PYR increased their firing rate during blue light, and no PYR decreased their firing rates during red light. These observations are consistent with inhibitory synaptic connections between the optically manipulated PV cells and the putative PYR

A key target application of dual-color optogenetics is to allow the bidirectional control of the same neuron. For that to happen, several conditions must be fulfilled at the same time. First, two opsins, each sensitive to a different wavelength and having a distinct effect on the membrane potential must be expressed within the same cell. Second, the two wavelengths must be applied to that cell. Indeed, some of the units that modulated their activity during blue and red light application also increased their spike rate during blue light and decreased during red light (9/74 of the INT; 12%). Most (but not all) of these optically doubly-verified PV cells had waveforms and autocorrelation histograms consistent with their classification as interneurons (Fig. 8A, 8B). Furthermore, almost half (4/9) of these units also displayed inhibitory connections with other simultaneously-recorded units (Fig. 8B), and were thus triply-verified inhibitory cells. In summary, the dual-color unit technology allows direct bidirectional control of individual well-isolated PV cells during free movement of small rodents.

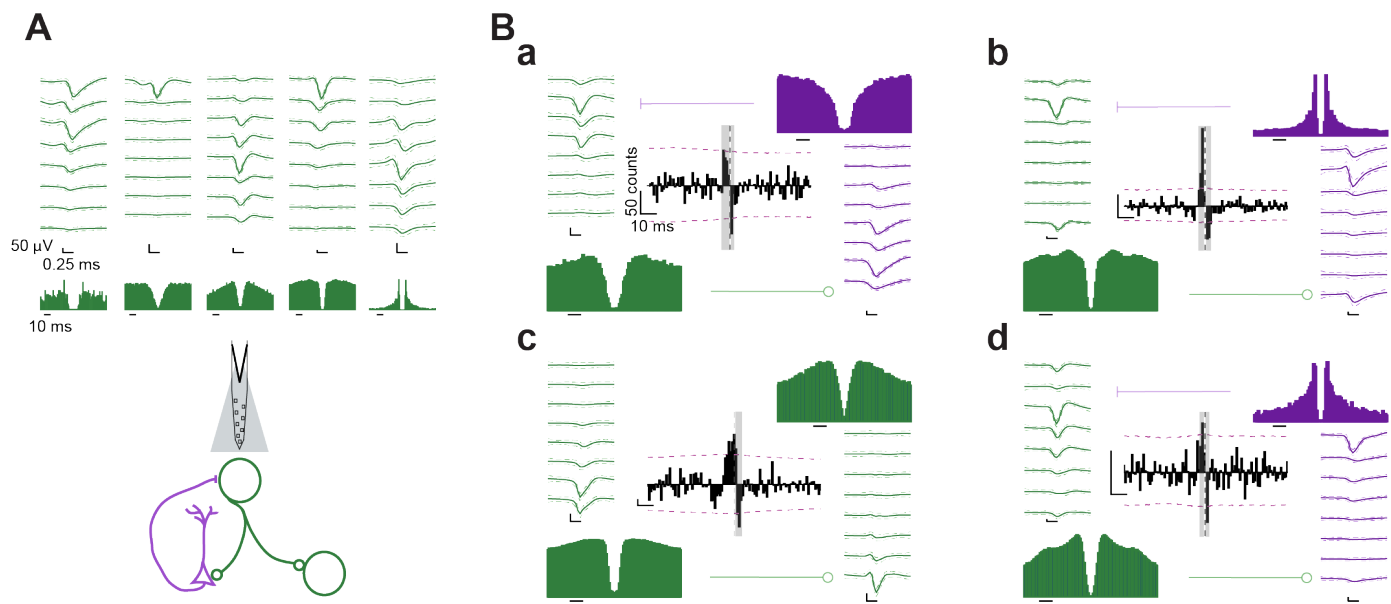

**Fig. 8. Properties of bidirectionaly controlled PV cells.** (**A**) Wideband (0.1–7,500 Hz) spike waveforms and autocorrelation histograms of five of the bidirectionaly controlled PV interneurons (top). Left-most two units are from neocortex, others from hippocampus. A cartoon of a putative minimal network that involves a bidirectionaly controlled PV interneuron (bottom). (**Ba**) Cross-correlation histogram (CCH; center) shows inhibition and reciprocal excitation between a bidirectionaly controlled neocortical PV cell (left) and a putative PYR (right). Monosynaptic connections (highlighted in grey) are determined by crossing of the 99.9% global confidence bands (purple dashed lines). (**Bb, Bd**) Same as **Ba** for a hippocampal network. (**Bc**) Inhibition of a hippocampal putative PV interneuron (right) by a bidirectionaly controlled PV cell (left).

### C. Dual-color probes allow prolonged recordings free from electromagnetic and thermal artifacts

Shielded LED-based diode-probes do not induce EMI, allowing clean wide-band electrophysiological recordings in behaving rodents [15], [38], [43], [44]. However, LD-based optoelectronic devices in which injection LDs were placed directly on the probe and coupled via GRIN lenses to integrated waveguides induced EMI artifacts of two types: transient (at the onset and/or offset of each pulse), and "DC artifacts" (during the pulses themselves; [16], [17]). Visual inspection of the data recorded during both blue and red illumination in the dual-color mice revealed clean wide-band traces (Fig. 6Ba, Ca). While results may differ for electrodes with different impedances, these observations indicate that at the power levels used in these experiments, the shielding and grounding measures used to minimize EMI were effective.

Because after probe implantation, the LDs are ~30 mm away from the skull, brain heating is not a concern with the dual color probes. However, light output of LDs decreases with increased temperature. Although bench testing showed that the units do not heat above 31 °C even after prolonged illumination (Fig. 4I), ventilation within the on-head Faraday cage is limited and device temperature may increase. We therefore monitored the temperature of the LD metal can continuously on head in one mouse. We found that heating did not exceed 7 °C even during prolonged illumination (maximal current, 29 mA; pulse duration, 50 ms; duty cycle, 0.6%; Fig. 9A, right). While we could not make direct measurements of light output after implantation, we measured the effect of light on optically-responsive units. We found that while device temperature gradually increased, the induced firing rate did not decrease (Fig. 9A, center). Similar results were obtained with duty cycles up to 50% (Fig. 9B). In summary, the power drawn by the dual-color units and used in freely-moving mice was sufficiently-low to prevent EMI and maintain low device temperature, while the light output was sufficiently high for effective activation and silencing of individual PV cells.

## IV. DISCUSSION

The design presented in this work provides a versatile, light weight platform for *in vivo*, head-mounted, dual-color optical stimulation. In essence, we compressed a full optical table, typically weighing hundreds of pounds, into an implantable unit weighing under one gram. Multiple units were mounted on each multi-electrode array and implanted in the mouse brain, providing dual-color control at several distinct locations. Up to four units can be implanted on a single mouse, and clearly more on larger animals. The dual-color unit delivers sufficient light for optogenetic activation and silencing while using low power LDs and low currents: in animal experiments, peak currents of 50 mA were used, without generating excessive heat or inducing artifacts in the electrophysiological records.

Using the dual-color units, an individual PV cell could be activated and silenced. This is not the first time that the same neuron is both activated and silenced optically: previous demonstrations were in brain slices [8], [26], in a head-fixed mouse [16], and in a freely-moving rat [15]. However, the previous *in vivo* experiments used opsins (ChR2 and Halo3, or ChR2 and Arch3) under the CAG and/or CaMKII promoters, leading to expression in excitatory cells. In such cases, a non-illuminated/opsin-free recorded neuron may inherit a rate increase/decrease via glutamatergic connections from directly-illuminated, opsin-expressing pre-synaptic neurons. In contrast, when the opsin-expressing cell is inhibitory, synaptic effects are reversed and any rate increase following blue light illumination stems from direct illumination (with the caveat of disynaptic dis-inhibition). Thus the present work shows conclusively, for the first time, that the same neuron can be both activated and silenced in a freely-moving animal.

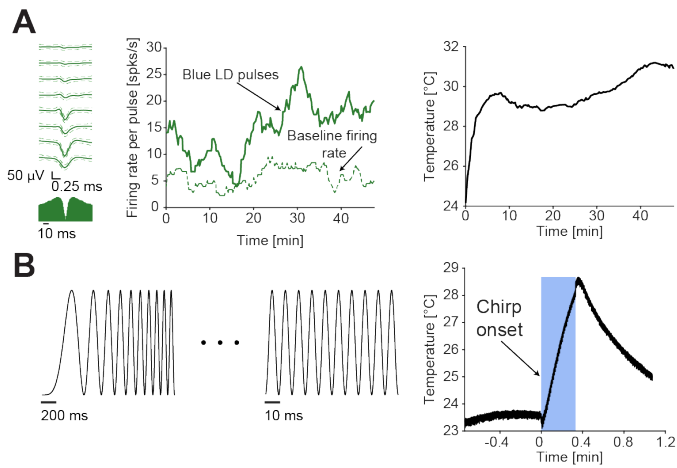

**Fig. 9. Device heating does not modify the effect of light on the neuronal response.** **(A)** Wideband (0.1–7,500 Hz) spike waveforms and autocorrelation histograms of an optically verified hippocampal PV interneuron (left). The PV cell firing rate per pulse (center) was calculated for each of 150 blue light pulses (50 ms, 29 mA, 0.6% duty cycle; green line) and 150 epochs with no stimulation intermingled between the stimulation pulses (50 ms, no driving current, dashed green line). Time is measured from the onset of the first pulse. The laser can temperature was monitored continuously on the mouse head. Heating did not exceed 7 °C (right). **(B)** An ascending chirp ($f_0 = 0$, $f_{end} = 100\ Hz$, 20 s ; left) was used for a high duty cycle (50%) stimulation. Heating did not exceed 6 °C (right; time is measured form chirp onset).

The dual-color units have multiple potential applications in freely-moving animals. First, bidirectional control of spike rate can be used to control spike timing [26], [15]. Second, intermingled neuronal populations with distinct genetic markers can be targeted [17], [5]. Third, some opsins ("step-function opsins"; [45], [31]) can be activated using one wavelength and deactivated using another wavelength, and dual-color units can be used for multi-site use of these opsins in freely-moving animals.

The main disadvantage of the presented approach is assembly duration: assembling a single dual-color optical unit requires about six hours of work. A second limitation is the size of the unit: although lighter than optical tables by four or five orders of magnitude, each unit weighs over 0.5 g. Thus, the number of units that can be placed on a small animal is limited to a handful. A third limiting factor is the percentage of PV interneurons with double expression of both CaMKII and Jaws. Ideally, all blue light-responsive units would also be red light-responsive. This can be achieved using a transgene consisting of linked ChR2 and Jaws proteins, e.g. using the P2 linker, as previously proposed [8].

The procedure described herein may be used to combine wavelengths other than blue and red. For instance, a red LD may be combined with a green LD to activate Archeorhodopsin [27] or a violet LD to activate ChR2 with minimal Arch activation. Ideally, all dual-color optical units should use TO38 can LDs to minimize device dimensions. For diodes emitting violet or near-green light, the same 550 nm mirror used here can be employed. If other colors – for instance, green (520 nm) and violet/blue diodes (400-450 nm) – are to be combined, a different dichroic mirror must be used (e.g. a 500 nm shortpass; Techspec #69-202, Edmund Optics). However, the assembly process would remain exactly the same.

The miniature dual-diode optical assembly described in this work can be used for completely different neuroscience applications such as fluorescence imaging of multiple deep sites in freely-moving small animals. For example, calcium imaging requires one wavelength for activation of the fluorophore and another for reading the calcium-dependent fluorescence [46]. *In vivo* calcium imaging activation is typically achieved using two-photon microscopy [47] - [49]. Currently, calcium imaging in freely moving animals is either limited to a single site due to physical limitations [50], [51] or requires optic fiber connections that restrict animal movement [52] - [54], [13]. Using the procedure described in this work, calcium imaging can be achieved simply by replacing one of the LDs with a miniature photodiode to record light picked up by the fiber, and choosing a dichroic mirror with a suitable wavelength. In such an application, the same multimode fiber will be used to activate the fluorescent protein using one wavelength (for example 485 nm when using GCamp6 [55]) and to receive the emitted light at a different wavelength (515 nm). This will allow recording fluorescence signals from multiple deep sites (e.g. brain regions) of small behaving animals with only electrical connections (possibly wireless) to the animal.

Finally, the dual-color approach could be extended to a triple-color approach by including another dichroic parallel to the first one, exactly as done in bench-top multi-color laser systems. This may allow independent control of three intermingled neuronal populations, bidirectional control of one population while activation or silencing another population, and other combinations. Presently, the activation spectra of available opsins overlap considerably in the visible range, but this bottleneck may be circumvented in the IR or UV ranges, and/or solved by the future development of narrow-band opsins.

## CONCLUSION

We developed a light weight platform for multi-site, dual-color optical stimulation combined with high-density extracellular recordings in freely-moving mice. Each dual-color unit can emit multiple wavelengths at the same point in the brain without excessive heating, allowing concurrent bidirectional optical control and clean multi-site neuronal recordings. Devices were used for activating and silencing PV cells in neocortex and hippocampus of freely-moving mice. The technology can be used for controlling spatially intermingled neurons that have distinct genetic profiles, and for controlling spike timing of cortical neurons during cognitive tasks.

conceived and supervised the project, designed the dual-color optical unit, built optoelectronic probes, implanted animals, analyzed data, and wrote the manuscript.